# Pseudospin-mediated broadband hypersonic phonon transport beyond localized Berry curvature

Byunggi Kim[1,2,*,†], Michele Diego[3,†], Yohei Ogawara[3], Matteo Pirro[3], Hyuga Oka[3], Ryo Sasaki[4], Satoshi Iwamoto[3,5], and Masahiro Nomura[3,*]

* Corresponding authors: kim.b.4dbc@m.isct.ac.jp, nomura@iis.u-tokyo.ac.jp
[1]Department of Mechanical Engineering, Institute of Science Tokyo, Tokyo 152-8550, Japan
[2]Space Innovation Visionary Initiative, Institute of Science Tokyo, Tokyo 152-8550, Japan
[3]Institute of Industrial Science, The University of Tokyo, Tokyo 153-8505, Japan
[4]RIKEN Center for Quantum Computing (RQC), Saitama 351-0198, Japan
[5]Research Center for Advanced Science and Technology, The University of Tokyo, Tokyo 153-8505, Japan
[†]These authors are equally contributed.

## ABSTRACT

Phononic nanostructures offer compact and highly confined control of elastic waves for signal transport, filtering, and quantum transduction. More recently, topological phononic structures have been explored as a route to robust waveguiding around defects and sharp bends, offering a promising basis for scalable and reliable phononic circuitry. However, how robust transport can persist as conventional localized Berry-curvature signatures diminish with increasing structural perturbation remains elusive. Here, we demonstrate bend-resilient hypersonic phonon transport beyond the conventional valley Hall framework. Using an engineered propeller phononic crystal operating at 1.9 GHz, we continuously increase the geometric perturbation to gradually weaken the valley topology while enhancing modal pseudospin character. Strikingly, as the Berry curvature progressively delocalizes from valleys, the complete elastic bandgap broadens to a normalized width of 34% and the interface modes become more strongly confined. Two-port electromechanical measurements on a thin-film lithium niobate-on-SOI platform experimentally confirm broadband interface-mode transport with no measurable bend-induced loss, despite the absence of well-localized valley-topology signatures. Our findings establish modal pseudospin engineering as a powerful design principle for robust acoustic guiding beyond conventional topological constraints, paving the way for broadband hypersonic phononic circuitry for classical and quantum information processing.

## INTRODUCTION

Topological phases of matter revealed that wave transport can be governed not only by local material properties, but also by global geometric characteristics of the underlying band structure[1–3]. Initially established for electrons in condensed-matter systems, this principle was subsequently extended to classical waves, for which artificial periodic structures provide direct control over symmetry, coupling and dimensionality. Photonic crystals provided an early platform for emulating quantum Hall, quantum spin Hall and valley Hall physics with electromagnetic waves, including edge states capable of propagating around structural irregularities with suppressed backscattering[4–9]. These concepts were later transferred to acoustic and elastic systems, giving rise to topological phononics[10–13]. By structuring solids into phononic crystals (PnCs), electronic notions such as pseudospin, valley degree of freedom, Berry curvature and bulk–boundary correspondence can be encoded into collective mechanical modes. This development has enabled unconventional manipulation of sound and elastic waves, including guided transport along domain walls, routing around sharp bends and selective control of mechanical energy across length scales ranging from macroscopic acoustic structures to on-chip nanomechanical systems[14–17].

Extending topological phononics to hypersonic frequencies is particularly attractive because gigahertz mechanical waves can interface with radio-frequency electronics, photonic and solid-state quantum systems, holding potential for acoustic quantum computing[14,18–22]. Recent experiments have demonstrated valley- and pseudospin-dependent elastic-wave transport in chip-scale phononic structures, including gigahertz edge states that remain transmissive across disorder and sharp corners[14,17,23–25]. However, hypersonic operation also exposes limitations that are less restrictive in macroscopic acoustic systems. The vectorial nature of elastic waves produces multiple interacting polarizations and dense mechanical bands, making it difficult to open a complete bandgap while retaining an isolated interface mode[15,25]. Moreover, the phononic analogy to conventional topological systems is clearest in the weakly perturbed, small-bandgap regime close to the original symmetry-protected degeneracies[12,26]. As the perturbation and bandgap increase, the topological description can become less accurate, raising the question of how robust transport persists beyond this regime[26,27]. Consequently, experimentally realized hypersonic topological PnCs have generally operated within relatively narrow spectral windows[14,15,17]. This contrasts with conventional, non-topological PnCs, in which geometric and material optimization can generate much wider complete bandgaps[28–30]. More fundamentally, the physical origin of bend-resilient transport in reciprocal valley systems remains unsettled[27,31,32]. Unlike a Chern insulator with strictly unidirectional edge states, a valley Hall system contains counterpropagating modes and is protected only insofar as perturbations suppress intervalley scattering or preserve the relevant modal symmetry. Distinguishing the contributions of band topology, pseudospin polarization and ordinary modal confinement is thus essential for developing broadband hypersonic phononic circuitry.

Here, we demonstrate pseudospin-mediated hypersonic phonon transport that remains robust as the conventional valley-topological description breaks down with increasing structural perturbation. We realize a hypersonic propeller PnC where the geometric perturbation can be scaled continuously. This approach allows us to deliberately collapse the isolated two-band valley description. As the perturbation increases, the system opens a complete bandgap centered at 1.9 GHz, achieving a normalized width of 34% that rivals conventional non-topological phononic-shielding crystals while further enhancing modal pseudospin character. Crucially, we demonstrate both numerically and experimentally that the resulting interface modes retain high, bend-resilient transmission through sharp, sub-micrometer corners, even as the localized valley-Berry-curvature signatures become delocalized. By tracking the divergent evolution of Berry curvature and in-plane energy-flow moment, we show that pseudospin-mediated robust phonon routing can persist beyond the regime of well-defined valley topology. Our work bridges the gap between narrow-band topological transport and wide-band phononic shielding, offering a versatile platform for broadband gigahertz phononic signal processing.

## RESULTS

### Design of propeller PnC

We designed quasi-two-dimensional periodic nanostructures on a silicon membrane with broken mirror symmetry relative to the membrane plane. Figure 1a shows a triangular PnC shaped as a "propeller" with a lattice constant $a$ = 1.65 μm. The hexagon of red-dashed line is a unit cell, where its tilted view is shown in the left panel of the inset with the thickness $t$ = 220 nm. The unit cell consists of a circular nanowire with anchored three masses, thereby having the $C_{3v}$ symmetry. The circular nanowires have a gradually varying width around each mass. By perturbing the width of the circular nanowires, we break the mirror symmetry. From the non-perturbated structure (grey structure, the right panel of **Fig. 1a** inset), we rotated the geometric phase of the circular nanowires to make width of the nanowire around each mass becomes non-symmetrical (sky blue structure, the right panel of **Fig. 1a** inset). We define the geometric perturbation as the rotation angle $\theta$ of the circular nanowires (See **Supplementary Information** for design details).

We calculated the dispersion of the propeller PnC using the finite-element method. **Figures 1b** and **1c** show the three-dimensional band structure in the reciprocal space, which is 20% larger than the first Brillouin zone. Without the perturbation ($\theta$ = 0°), the Dirac cones-like band structures formed at 1.9 GHz. The positions of the Dirac cones in the momentum space were near the K points, which establish the valley degree of freedom required for a phononic analogue of valley Hall effect. The positions of the Dirac cones are slightly shifted from the K points due to the material anisotropy[33]. As the perturbation increases, the bandgap gradually widens, reaching a complete bandgap of 34% relative to its center frequency at $\theta$ = ±18°. **Figure 1d** shows the $\theta$-dependent bandgap and the mode shapes corresponding to the upper and lower bands for $\theta$ = ±18°. The mode could be characterized by the direction of the energy flux along circular nanowires: Clockwise for the red line and counter clockwise for the blue line. The inversion of the modal characters across $\theta$ = 0° reflects a reversal of the symmetry-breaking perturbation[34,35]. This modal inversion is a typical characteristic of valley Hall systems, in which reversing the symmetry-breaking perturbation changes the sign of the effective Dirac mass at the K and K′ valleys[35,36]. An interface between domains with opposite mass terms can support valley-polarized edge states, whose backscattering is suppressed as long as intervalley mixing remains weak. The associated Berry curvature and modal pseudospin are examined in the following subsection.

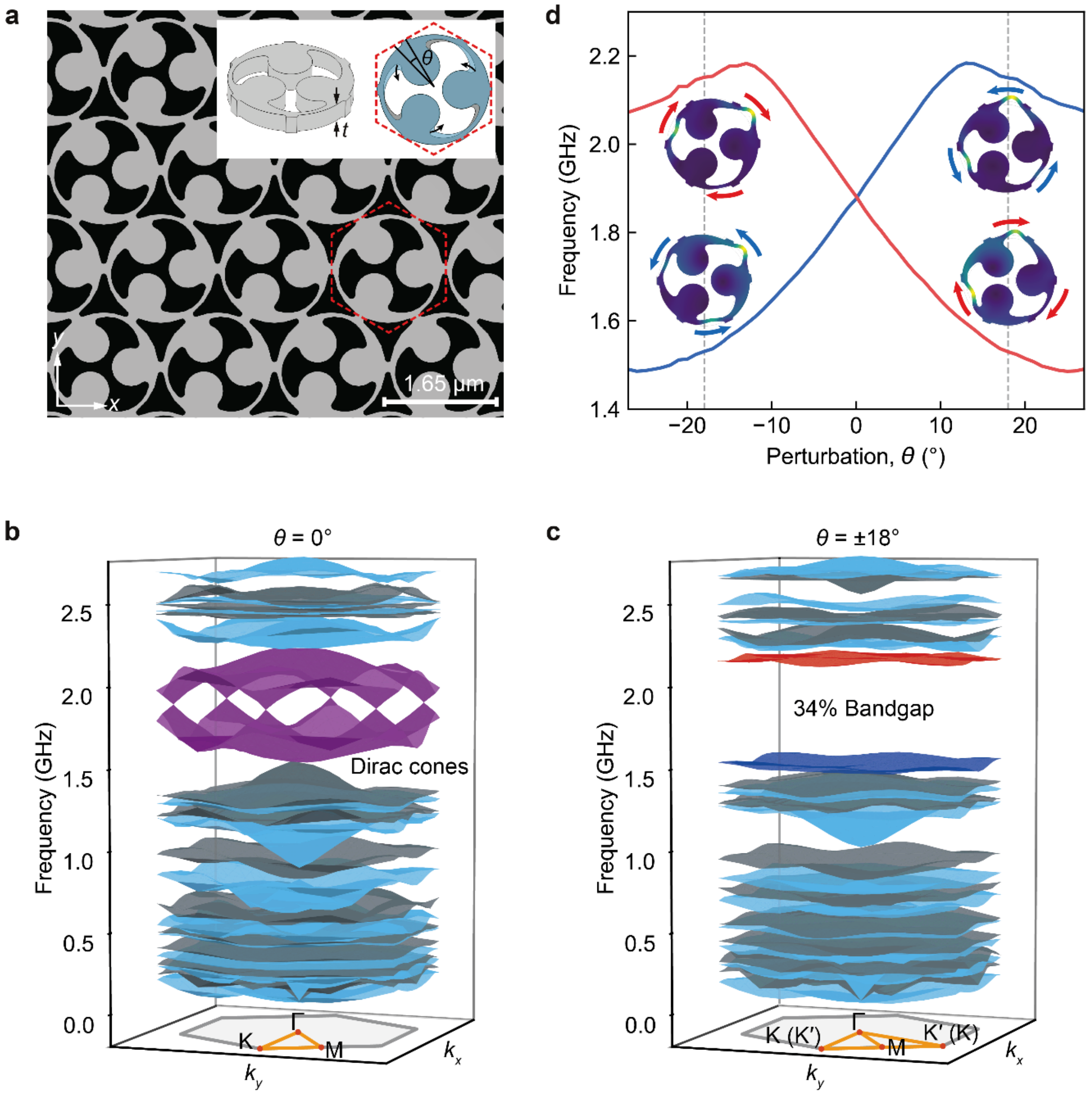

**Fig. 1 | Propeller PnC and its band structures. a,** Design of the propeller PnC. Grey and black regions denote silicon and vacuum, respectively. The inset shows tilted and top views of the unit cell. The unit cell, consisting of a circular nanowire connecting three enlarged masses, is patterned in a silicon membrane of thickness $t = 220$ nm. The $C_{3v}$ symmetry is broken by varying the geometric phase $\theta$ of the circular nanowire. **b**, **c**, Simulated three-dimensional band structures for $\theta = 0°$ and $|\theta| = 18°$, respectively. The first Brillouin zone is projected onto the $k_x$–$k_y$ plane. At $\theta = 0°$, Dirac cones appear at the K and K′ points. Breaking the mirror symmetry by introducing a finite θ lifts the Dirac degeneracy and opens a complete phononic bandgap with a relative width of 34% at $|\theta| = 18°$. **d**, Evolution of the two bands forming the Dirac cone as a function of $\theta$, showing band inversion across θ = 0°. Insets show representative eigenmodes at θ = −18° and +18°, with arrows indicating the opposite circulation of the local energy flux associated with the two modes.

**Berry curvature and modal pseudospin**

We calculated the Berry curvatures of the upper and lower bands bordering the bandgap using the Fukui–Hatsugai–Suzuki formalism[37,38] (**Methods**). **Figure 2a** shows the Berry curvatures evaluated along the momentum-space path indicated in the inset for different perturbation angles. At a small perturbation angle ($\theta = 4°$), the Berry curvatures are strongly localized near the K and K′ points and exhibit opposite signs for the upper and lower bands. Such valley-localized Berry curvature is characteristic of a regime capable of supporting valley-selective transport of elastic waves. As the perturbation increases, the Berry-curvature distributions delocalize across momentum space and, from $\theta = 8°$, lose the approximate upper–lower-band symmetry observed at smaller perturbations. Their delocalization indicates the breakdown of the isolated two-band valley description, suggesting an increasing influence of adjacent bands.

Pseudospin-like modal textures can also play an important role in the directional transport of elastic waves in valley Hall phononic systems. To quantify the modal pseudospin, we calculated the in-plane rotational energy-flow. The in-plane energy-flow moment was normalized by the magnitude of the in-plane energy flux:

$$C_z = \frac{\int [(y-y_c)I_x - (x-x_c)I_y]dv}{a\int \sqrt{I_x^2+I_y^2}dv} \quad (1)$$

where $I_x$ and $I_y$ denote $x$- and $y$-components of the mechanical energy flux, respectively. The center of rotation, $(x_c, y_c)$, was set at the center of the unit cell. This dimensionless quantity represents the circulation of the in-plane energy flow, with its sign indicating the direction of modal circulation. In **Fig. 2b**, the upper and lower bands carry the angular momenta of opposite signs, while their magnitudes are substantially larger at the stronger perturbation of $\theta = 18°$. Notably, although the K and K′ valleys are no longer well isolated at $\theta = 18°$, the modal pseudospin is enhanced. This contrast suggests that resilient wave transport may persist through pseudospin-selective modal dynamics even when the valley-based topological description becomes ill-defined, raising the possibility of pseudospin-mediated bend-resilient transport without a well-defined valley topological invariant.

**Figure 2c** benchmarks the normalized bandgap size, defined relative to the bandgap center frequency, against previously reported experimental works over 100 MHz[14,15,17,24,25,29,39,40]. The valley Hall phononic systems[14,15,17,24,25,39] have generally exhibited relatively narrow bandgaps, which may constrain their practical implementation as phononic mirrors. By contrast, wide-bandgap PnCs based on strain localization have enabled effective phononic shielding[40–42]. Although the normalized bandgap of the propeller PnC is slightly smaller than those of the widest reported PnCs, it combines a comparatively broad bandgap with pronounced pseudospin character. The proposed platform therefore offers a route bridging valley Hall-mediated transport at weak perturbations and pseudospin-dominated transport at stronger perturbations.

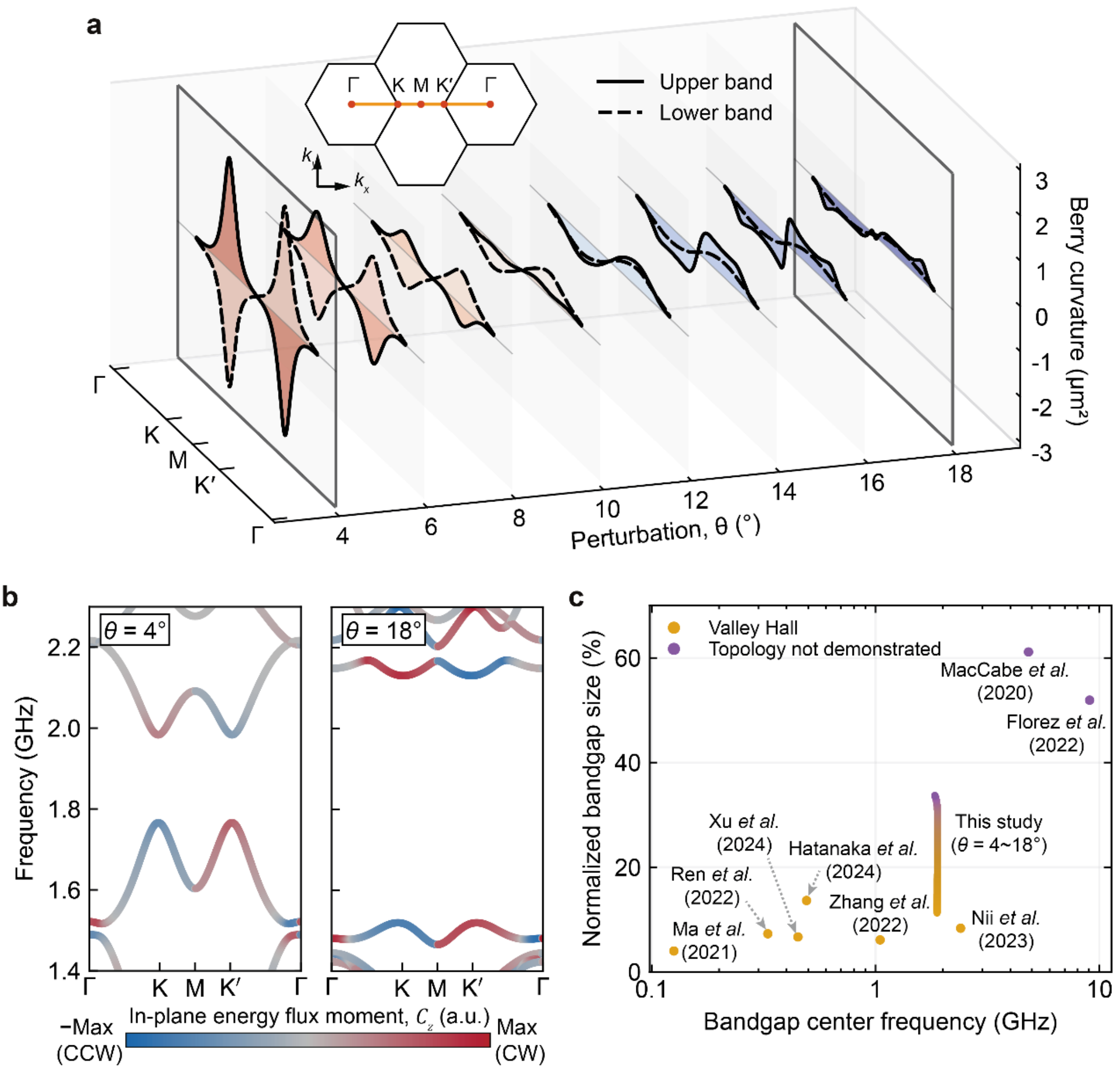

**Fig. 2 | Evolution of the topological characteristics with increasing geometric perturbation. a,** Evolution of the Berry curvature of the upper and lower bands forming the bandgap as the geometric phase $\theta$ is increased from 4° to 18°. The Berry curvature is strongly localized around the K and K′ valleys for a small perturbation, whereas it gradually delocalizes and becomes increasingly asymmetric between the upper and lower bands at larger $\theta$. Solid and dashed lines denote the upper and lower bands, respectively. The inset shows the first Brillouin zones and the reciprocal-space path Γ–K–M–K′. **b**, Band structures calculated at $\theta$ = 4° and 18°, with the color indicating the normalized in-plane energy-flow circulation (pseudospin) of each eigenmode. At $\theta$ = 4°, the pseudospin character is primarily associated with the vicinity of the valleys, whereas at $\theta$ = 18° pronounced opposite spin characters extend over a broader region of the bands. **c**, Comparison of the center frequency-normalized bandgap and operating frequency of previously reported topological phononic systems over 100 MHz. Symbols indicate valley-Hall phononic systems (orange) and phononic structures for which topology has not been demonstrated (purple). The highlighted range indicates the present structure for $\theta$ = 4°~18°, reaching a complete bandgap of 34% at the largest perturbation.

**Transport along domain walls**

Domain walls between PnCs can support localized interface modes. When the adjoining domains possess distinct topological invariants, these modes emerge as topological edge states capable of bend-resilient transport with suppressed intervalley backscattering. Recent studies in photonic systems have shown that bend-resilient transport does not necessarily require a well-defined valley topology, and can instead originate from the modal characteristics or domain-wall configuration[43–45]. This raises the question of whether an analogous mechanism can sustain robust transport for elastic waves beyond the conventional valley-topological regime. To distinguish the roles of valley topology and pseudospin in phononic transport, we therefore compare three domain-wall configurations providing a topological benchmark, a pseudospin-preserving non-topological case, and a conventional-defect case.

**Figure 3** presents the calculated dispersions and wave propagation for three domain-wall configurations. As illustrated in the left panels of **Figs. 3a–c**, the three interfaces are designed to separate the contributions of valley topology and pseudospin. The A–B interface at $\theta$ = ±4° retains both valley-topological and opposite pseudospin textures and therefore serves as the topological benchmark (**Fig. 3a**). Increasing the perturbation to $\theta$ = ±18° yields an A–B interface for which a conventional valley-topological invariant is no longer well defined, while the opposite pseudospin textures remain (**Fig. 3b**). Finally, the A–A′ interface, formed by two $\theta$ = 18° domains related by a 60° rotation, provides non-topological defect modes without the direction–pseudospin locking (**Fig. 3c**). The right panels of **Figs. 3a–c** show the supercell dispersions of the corresponding domain walls. Purple and green denote predominantly in-plane and out-of-plane modes, respectively. For the A–B domain walls ($\theta$ = ±4° and $\theta$ = ±18°), the in-plane interface modes exhibit opposite pseudospin textures, characterized by counter-rotating in-plane energy fluxes. The enlarged bandgap at $\theta$ = ±18° accompanies stronger spatial confinement of the elastic energy to the domain wall, consistent with drastic attenuation into the surrounding PnC. These interface modes support bend-resilient propagation through the corresponding waveguides, as shown in **Figs. 3d** and **3e**. By contrast, the domain A–A′ interface in **Fig. 3c** supports multiple defect modes distributed across the bandgap, which do not inherently exhibit bend resilience, as demonstrated in **Fig. 3f**.

**Figure 3g** shows the simulated transmission spectra through the Z-shaped waveguides, normalized by those of PnCs without domain walls. The interface modes formed at the $\theta$ = ±4° and $\theta$ = ±18° domain walls exhibit significant transmission over their respective bandgaps. In contrast, the transmission through the A–A′ interface ($\theta$ = 18°) remains low over most of the bandgap spectra. **Figure 3h** compares the integrated transmission through the straight (I-shaped) and bent (Z-shaped) waveguides for the pseudospin modes at the domain A–B interfaces ($\theta$ = ±4°, ±18°) and the conventional defect modes at the domain A–A′ interface ($\theta$ = 18°). For the pseudospin-paired interface modes, the integrated transmissions through the I- and Z-shaped waveguides are comparable, demonstrating resilience against sharp bends, while the broad bandgap formed at the $\theta$ = ±18° supports the larger overall transmission. The results for $\theta$ = ±18° therefore show that pseudospin-selective interface transport can remain bend-resilient even when a well-defined valley topological invariant is absent. The conventional defect modes at the A–A′ interface ($\theta$ = 18°), however, exhibit reduced transmission even through the straight waveguide, possibly because of mode mismatch and scattering among the multiple coexisting defect modes. Building on analogous observations in photonic systems, the present phononic platform reveals an additional consequence of strong perturbation: substantial broadening of the complete elastic bandgap while preserving pseudospin-mediated bend resilience.

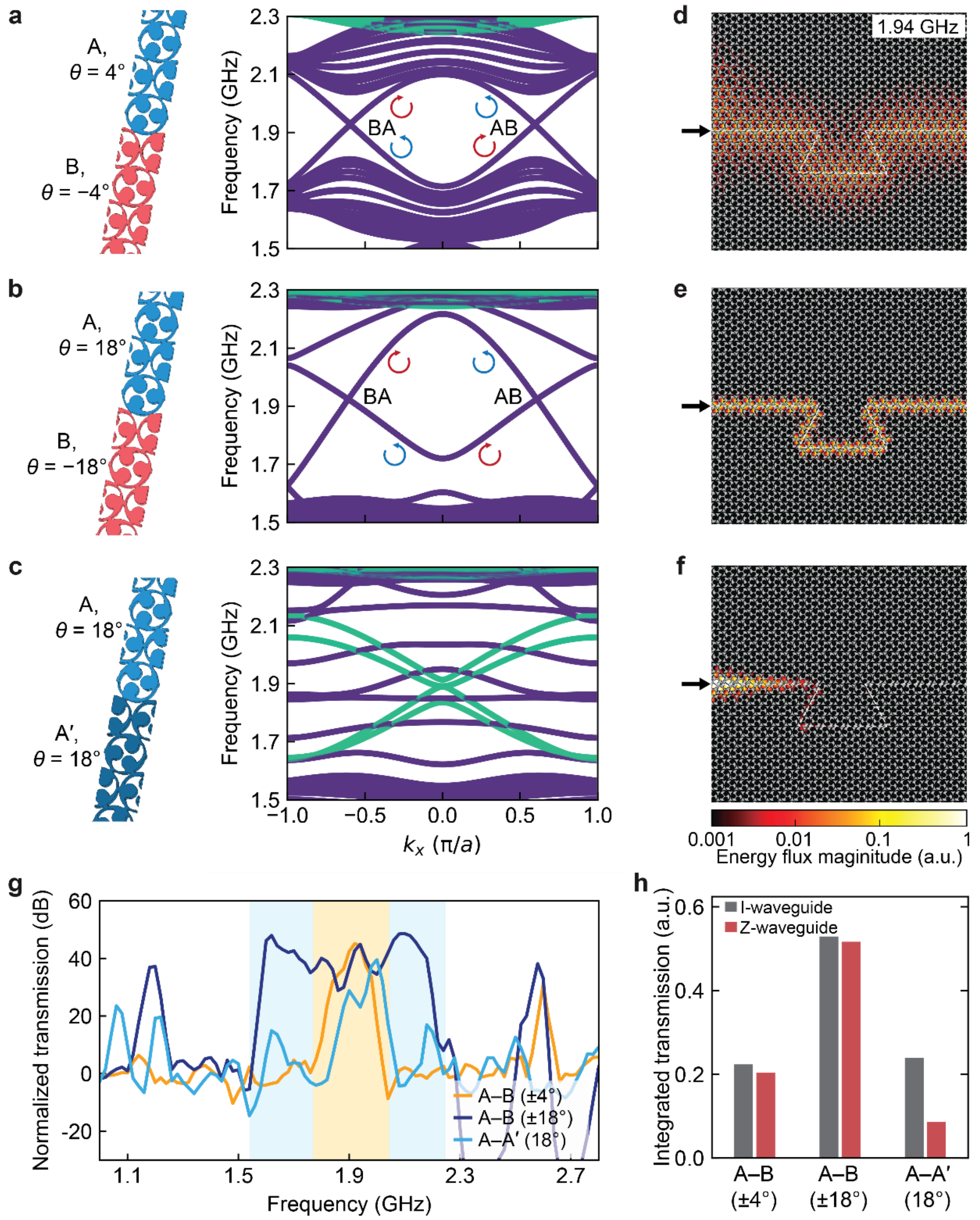

**Fig. 3 | Pseudospin-mediated interface modes and robust phonon transport. a–c,** Interface configurations (left) and corresponding supercell band structures (right) for the domain walls formed by $\theta = \pm 4°$ (**a**) and $\pm 18°$ (**b**), and a 60° lattice rotation at $\theta = 18°$ (**c**). The A and B domains are related by reversal of the geometric perturbation, whereas A′ is obtained by a 60° rotation of domain A. For the A–B interfaces, pairs of counter-propagating interface modes appear within the bulk bandgap. Purple and green curves denote out-of-plane and in-plane modes, respectively. Circular arrows indicate the opposite pseudospin characters of the interface modes. **d–f,** Simulated elastic energy density distributions at 1.94 GHz, showing wave propagation through Z-shaped bent waveguide. The pseudospin-paired interfaces (d, e) exhibit robust, bend-resilient transmission, whereas the conventional defect modes in the rotated interface (f) suffer from severe backscattering and energy delocalization. **g,** Simulated transmission spectra for the A–B interfaces with $\theta = \pm 4°$ and $\pm 18°$ and the A–A′ interface at $\theta = 18°$. Shaded regions indicate the corresponding bulk bandgaps. **h,** Comparison of the band-integrated transmission through I-shaped and Z-shaped waveguides for the different interface configurations.

**Figure 4a** schematically illustrates the two-port measurement used to characterize elastic-wave transmission through the PnC devices. A microwave signal from port 1 of a vector network analyzer (VNA) is converted into elastic waves through piezoelectric coupling at a chirped interdigital transducer (IDT) fabricated on thin-film lithium niobate (TFLN) bonded to a silicon-on-insulator (SOI) substrate. After propagating through the suspended silicon membrane containing the propeller PnC, the elastic waves are converted back into a microwave signal by the receiving IDT and detected at port 2 (**Methods**). Therefore, the measured $S_{21}$ parameter represents the overall electromechanical transmission of the device, including elastic-wave propagation through the PnC region. The purple line indicates the Z-shaped interface-waveguide path within the PnC, whose scanning electron microscopy image is shown in **Fig. 4b**.

The normalized transmission was obtained by normalizing the $S_{21}$ spectrum of each PnC device to that of an intact suspended silicon membrane. **Figure 4c** compares the normalized transmission spectra of four device configurations: a gapless PnC ($\theta = 0°$), a gapped PnC without an interface waveguide, and gapped PnCs incorporating straight I-shaped and bent Z-shaped waveguides formed between domains for $\theta = \pm 4°$ and $\theta = \pm 18°$. The shaded regions represent the standard deviations obtained from measurements of four nominally identical devices. Within the IDT operating bandwidth, defined by its full width at half maximum, the gapped PnC without an interface waveguide exhibits consistently lower transmission than the other configurations. This suppression is consistent with the phononic bandgap inhibiting elastic-wave propagation. Introducing the $\theta = \pm 4°$ and $\theta = \pm 18°$ domain walls produce a reproducible enhancement in transmission over the same frequency range, indicating that the interface modes provide a preferential propagation channel. Notably, the transmission spectra of the I- and Z-shaped waveguides remain comparable within the device-to-device variability, indicating no measurable loss from the sharp bends. Interestingly, the domain-wall devices of $\theta = \pm 4°$ exhibit slightly higher transmission than the devices of $\theta = \pm 18°$. This observation suggests that the measured port-to-port transmission may also depend on IDT-to-PnC coupling, possibly through differences in modal overlap between the IDT-launched elastic field and the interface modes. Together with the calculated pseudospin textures of the interface modes, these experimental results are consistent with pseudospin-mediated bend-resilient transport even in the strongly perturbed regime, where the conventional valley-topological description is no longer well defined ($\theta = \pm 18°$).

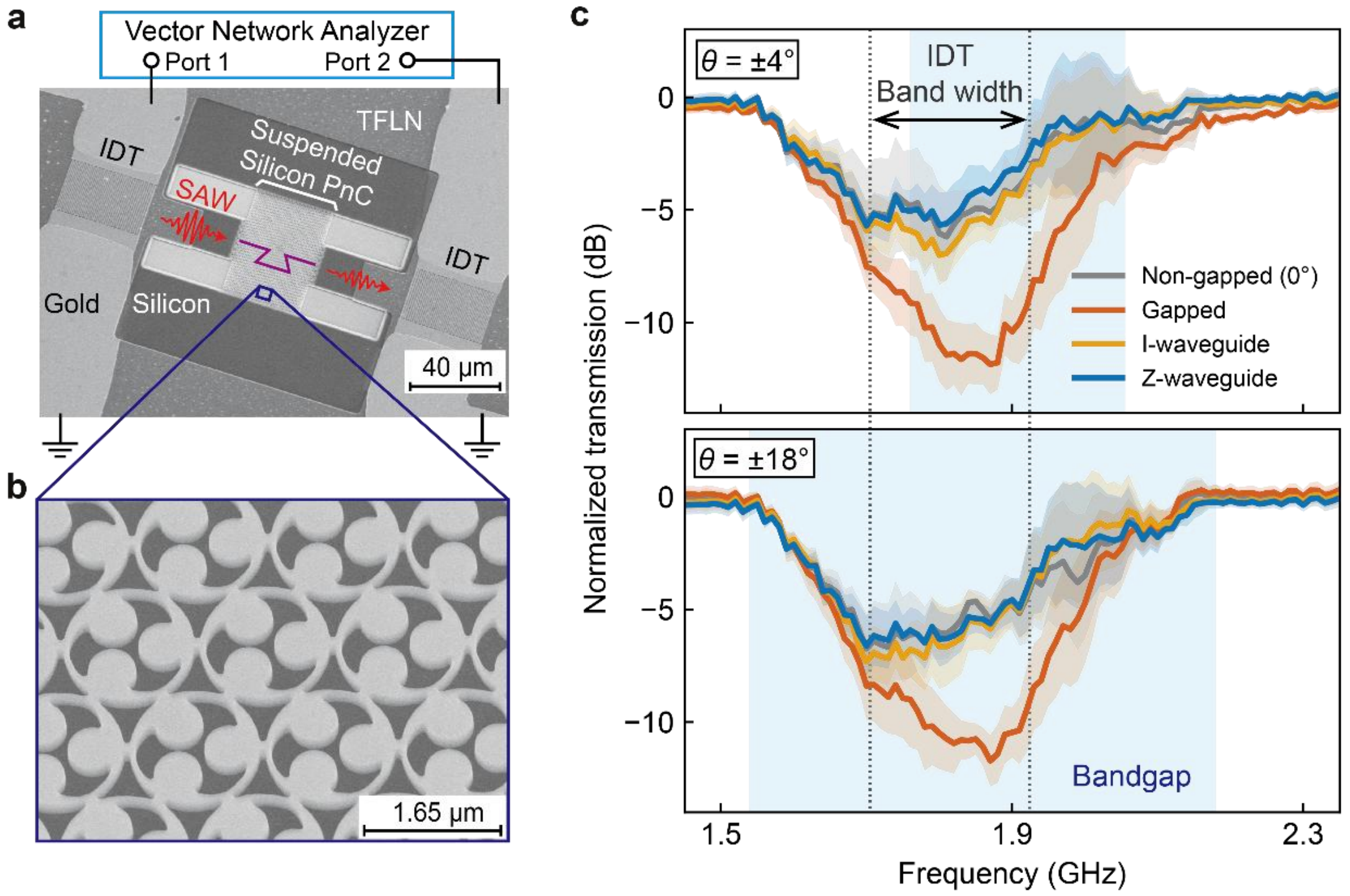

**Fig. 4 | On-chip experimental demonstration of broadband, bend-resilient hypersonic phonon transport. a,** Scanning electron microscope image of the fabricated TFLN-on-SOI device, in which input and output IDTs launch and detect SAWs across a suspended silicon PnC region. **b**, Enlarged scanning electron microscope image of the propeller PnC lattice. **c**, Normalized transmission spectra measured for the non-gapped structure ($\theta = 0°$), the gapped

structure without waveguides, and the gapped structure with I-shaped and Z-shaped waveguides for $\theta = \pm 4°$ (top) and $\theta = \pm 18°$ (bottom). The simulated complete bandgap is highlighted in blue, and the dotted lines indicate the IDT bandwidth. The interface waveguides restore transmission within the bandgap, in clear contrast to the gapped structure without an interface. Moreover, the comparable responses of the I-shaped and Z-shaped waveguides indicate that phonon transport is preserved even in the presence of sharp bends. Shaded envelopes represent the standard deviation across four nominally identical devices.

## DISCUSSION

Our results reveal an evolution in the physical mechanism underlying resilient hypersonic phonon transport as the structural perturbation is increased. At weak perturbations, band inversion is accompanied by Berry curvature localized near the K and K′ valleys, consistent with the conventional valley Hall picture, while the band-edge states also exhibit finite pseudospin character associated with opposite in-plane energy-flow moments. As the perturbation increases, these two characteristics evolve differently: the Berry curvature becomes delocalized and the isolated two-band valley description is no longer applicable, whereas the pseudospin character is retained and substantially enhanced. This contrasting evolution shows that valley topology and modal pseudospin can coexist in the weak-perturbation regime but need not remain correlated at stronger perturbations, identifying pseudospin as an independent degree of freedom for engineering elastic-wave transport.

Despite the absence of a well-defined valley topological invariant at large perturbations, domain walls between the $\theta = \pm 18°$ structures support the pseudospin-mediated interface modes with opposite pseudospin textures. Numerical simulations show that these modes propagate through sharp bends while retaining their modal character, in contrast to the multiple conventional defect modes supported by the rotated-cells interface. Experimentally, the domain-wall devices exhibit enhanced transmission relative to the gapped PnC without an interface channel. Moreover, the straight and sharply bent waveguides show comparable transmission within the device-to-device variability, demonstrating the absence of a measurable bend-induced transmission penalty. Together, these results establish pseudospin-mediated bend-resilient transport beyond the conventional valley Hall framework.

The propeller PnC further combines this transport functionality with a wide bandgap at hypersonic frequencies. Previous topological phononic systems have generally relied on relatively narrow bandgaps associated with weak symmetry-breaking perturbations, whereas wide-bandgap PnCs have primarily been developed for phononic shielding and thermal-phonon engineering without resilient interface transport. By uniting a broad phononic bandgap with pseudospin-paired interface modes, our platform bridges these previously distinct regimes. This approach expands the design space for resilient elastic-wave manipulation and offers a route toward broadband phononic routing, shielding, and integrated signal processing at gigahertz frequencies.

## METHODS

### Berry curvature calculation

First, the phononic band structures and eigenmodes of the propeller PnC were calculated using the finite-element method implemented in COMSOL Multiphysics 6.1, with Bloch–Floquet periodic boundary conditions applied to the unit-cell boundaries. Second, the Berry curvatures of the upper and lower bands bordering the bandgap were evaluated using the gauge-invariant discretized formulation of Fukui, Hatsugai and Suzuki. Complex inner products between eigenmodes at neighboring wavevectors were calculated in COMSOL. For each reciprocal-space plaquette defined by four neighboring wavevectors labelled 1–4 in cyclic order, the link variable between adjacent states was defined as

$$U_{mn} = \langle u_m | u_n \rangle / |\langle u_m | u_n \rangle|, \tag{2}$$

where $u_m$ and $u_n$ denote the complex displacement field of the eigenmodes at adjacent wavevectors. The Berry curvature associated with each plaquette was then calculated from the accumulated phase around the closed loop as

$$\Omega = \arg(U_{12}U_{23}U_{34}U_{41}) / (\Delta k_x \cdot \Delta k_y), \tag{3}$$

where $\Delta k_x \cdot \Delta k_y$ is the area of the reciprocal-space plaquette. The Berry curvature was evaluated along the momentum-space path shown in **Fig. 2a** for the upper and lower bands immediately above and below the bandgap, respectively, at each perturbation angle.

**Device fabrication**

The devices were fabricated on a TFLN-on-SOI platform, commercially available from NGK Insulator Ltd. Large-scale structures, including contact electrode pads, coplanar waveguides and etching windows on the TFLN, were defined by the ultraviolet photolithography, whereas fine features, including the IDT and PnC patterns, were defined by the electron-beam lithography. Gold electrodes were formed by metal deposition followed by lift-off. The TFLN layer was etched by argon-ion milling to expose the silicon layer. The PnC geometry was subsequently transferred into the silicon device layer by sulfur-hexafluoride reactive-ion etching. Finally, the buried $SiO_2$ layer underneath the PnC region was selectively removed by vapor-phase hydrofluoric-acid etching, releasing the silicon membrane and forming the suspended PnC structure.

**Phonon transport spectroscopy**

Phonon transport through the fabricated devices was characterized using two-port VNA (P5004B, Keysight) measurements. The input and output ports of the VNA were connected to the devices through the ground-signal-ground probes and coplanar waveguides terminating at chirped IDT comprising 35 pairs of fingers and exhibiting a 10% fractional 3-dB bandwidth. The transmitting IDT converted the applied microwave signal into elastic waves through piezoelectric coupling in the Y-cut TFLN, with the propagation direction of the Love-like mode is aligned along the X axis. The receiving IDT converted the transmitted elastic waves back into an electrical signal. The forward transmission coefficient, $S_{21}$, was measured as a function of frequency to characterize elastic-wave transmission through the suspended PnC region.

## ACKNOWLEDGEMENT

We thank Dr. Satomi Ishida and Mr. Hideo Otsuki for their technical support and fruitful discussions regarding argon-ion milling of the TFLN.

## AUTHOR CONTRIBUTIONS

B. K and Y. O. conceived the idea. B. K., M. D., S. I. and M. N. performed conceptualization. B. K. and Y. O. performed the simulations. B. K., M. D. and H. O. carried out the sample fabrication. B. K., M. P. and M. D. implemented the experimental setup. B. K., M. D. and R. S. conducted the formal analysis. B. K. and M. D. wrote the original manuscript. M. N., B. K. and M. D. acquired the fundings. M. N. supervised the project. All authors discussed and reviewed the manuscript.

## FUNDINGS

M. N. acknowledges support from JST ASPIRE (grant No. JPMJAP2316) and JSPS KAKENHI (grant No. 21H04635 and 24KF0027); B. K. acknowledges supports from the Asahi Glass Foundation and Research Foundation for Opto-Science and Technology; M. D. acknowledges supports from JSPS KAKENHI (25K17905). S. I acknowledges supports from JSPS KAKENHI (26K21743).

## CONFLICT OF INTEREST

The authors declare no conflict of interest.

## DATA AVAILABILITY STATEMENT

The data that supports the findings of this study are available on the reasonable request to the corresponding authors.